\documentclass[12pt,awide]{article}
\usepackage{latexsym,a4,a4wide,psfrag,here,
amssymb,amsmath,axodraw,color,ifthen,mathrsfs}
\usepackage[dvips]{graphicx}
\usepackage[mathcal]{eucal}
\usepackage{epsfig}

\def\mA{\ensuremath{\mathcal{A}}}

\def\mH{\ensuremath{\mathcal{H}}}
\def\mL{\ensuremath{\mathcal{L}}}

\def\mO{\ensuremath{\mathcal{O}}}

\def\mfL{\ensuremath{\bar{\mathfrak{L}}}}

\def\nn{\nonumber \\}
\def\no{\nonumber}

\def\fs{\; \; .}
\def\co{\; \; ,}
\def\con{\;\;, \nonumber \\}
\def\sem{\; \; ;}
\def\fs{\; \; .}
\def\co{\; \; ,}
\def\sem{\; \; ;}

\def\mln{\mbox{ln}}

\def\arctan{\mbox{arctan}}

\def\mpi{\ensuremath{m_{\pi}}}

\def\mk{\ensuremath{m_K}}
\def\mk{\ensuremath{m_K}}

\def\me{\ensuremath{m_{\eta}}}

\def\eps{\ensuremath{\varepsilon}}

\def\lp{\ensuremath{\mu_{\pi}}}
\def\lk{\ensuremath{\mu_{K}}}
\def\le{\ensuremath{\mu_{\eta}}}

\def\las{\tilde{\lambda}}

\def\cpt{\mbox{CHPT\,}}

\def\bdm{\begin{displaymath}}
\def\edm{\end{displaymath}}
\def\be{\begin{equation}}
\def\ee{\end{equation}}

\def\bea{\begin{eqnarray}}
\def\eea{\end{eqnarray}}
\def\bes{\begin{eqnarray*}}
\def\ees{\end{eqnarray*}}

\def\ba{\begin{array}}
\def\ea{\end{array}}
\def\bal{\begin{align}}
\def\eal{\end{align}}
\def\bspl{\begin{split}}
\def\espl{\end{split}}
\def\bml{\begin{multline}}
\def\eml{\end{multline}}

\def\um{\ensuremath{u_{\mu}}}

\def\uum{\ensuremath{u^{\mu}}}

\def\De{\ensuremath{\Delta}}

\def\cp{\ensuremath{\chi_{_{+}}}}

\def\u2{\ensuremath{u^{2}}}

\def\ltr{\langle}
\def\rtr{\rangle}

\def\phy{_{\mbox{\tiny{ph.}}}}

\def\ktpp{\ensuremath{K \to \pi \pi \,\,}}
\def\kztpp{\ensuremath{K^{0} \to \pi \pi \,\,}}
\def\ktp{\ensuremath{K^{0} \to \pi^{0}\,\,}}
\def\ktv{\ensuremath{K^{0} \to | 0 \rtr\,\,}}
\def\kkpp{\ensuremath{K^{0} \bar{K}^{0} \to \pi \pi \,\,}}

\def\bdm{\begin{displaymath}}
\def\edm{\end{displaymath}}

\renewcommand{\theequation}{\arabic{equation}}

\def\la{\langle}
\def\ra{\rangle}

\def\GeV{\mbox{GeV}}
\def\MeV{\mbox{MeV}}

\def\epe{\varepsilon'/\varepsilon}

\def\ho{$ \hbar $-order }

\definecolor{gray}{gray}{.5}
\definecolor{steelblue}{rgb}{0.153, 0.102, 0.255}
\definecolor{lightblue}{rgb}{0,0.2,0.5}

\renewcommand{\i}{{\rm i}}

\renewcommand{\theequation}{\arabic{section}.\arabic{equation}}

\begin{document}
\bibliographystyle{aipep}
\title{The chiral logs of the \ktpp amplitude}
\author{
 M.~B\"uchler$^{a}$ \vspace{1cm}\\
{\small
${}^a$ Department of Physics, University of Washington}\\
{\small Seattle, WA 98195-1560, U.S.A. }}
\maketitle
\begin{abstract}
I calculate the leading logarithmic contributions up to two-loop order of
the octet part of the \kztpp amplitude. This sector of the weak chiral Lagrangian is believed
to be the main source of the enhancement of the $I=0$ relative to the
$I=2$ \kztpp amplitude, the so-called $\De I = 1/2$ rule.
I discuss the procedure of chiral extrapolations of lattice data specific to \ktpp decays
and study the implication of the present calculation on these numerically. The latter reinforces
the fact that one has to expect a large enhancement of the $I=0$ part of the amplitude due to
re-scattering effects between the three mesons. 
\end{abstract}
\section{Introduction}
Chiral logs are introduced during the process of renormalization \cite{Li:1971vr}; 
In the framework of dimensional regularization, one has to introduce an energy scale $\nu$
to ensure the correct dimension of observables:
Terms like the one on the LHS of Eq. (\ref{eq:clo}), generated by loop integrals in $d = 4 - \eps$ dimensions, are re-expressed  in the $\nu$ independent  form on
the RHS \footnote{These terms are multiplied by geometrical factors and polynomials of integer powers in
the masses.}. It is necessary to adopt the energy scale $\nu$ to provide the proper dimension of the divergent and finite pieces in a well-defined way. Physically, the emerging chiral logs
can be associated with the infrared singularities when the masses of the theory approach zero, 
\bea
 \Big( \frac{m^{-\eps}}{\eps} \Big)^{n} \to \Big( \nu^{-\eps} \big( \frac{1}{\eps}
- \frac{1}{2} \mbox{ln}(\frac{m^{2}}{\nu^{2}}) \big) \Big)^{n} \co 
\label{eq:clo}
\eea
and can produce sizeable contributions to amplitudes. \\
Chiral logarithms are an important ingredient in a lattice determination of the \kztpp
amplitude: the lattice community commonly uses a standard method to relate the latter
to the \ktp and \ktv matrix element \cite{Bernard:1985wf}. However, this
approach cannot account for QCD interactions between the mesons, in particular the so called final state
interactions between the two pions. These effects, though, are expected (at least partially) to explain the
enhancement of the amplitude with the two pions in the $I=0$ state relative to the $I=2$ state
($\De = 1/2$ rule) as well as the disagreement of the value of $\epe$ between theory
and measurement. \cpt provides a tool to estimate the effects caused by these interactions,
and the chiral logs are a numerically important part thereof. \\
In two flavor \cpt, these contributions are commonly the dominant part of
the NNLO corrections, which is illustrated in the case of $\pi\pi$ scattering in \cite{Bijnens:1995yn}.  In the case of three flavor \cpt, the double log contributions are however not as prominent. Table \ref{t:b2l} shows the chiral corrections up to NNLO to the pion and Kaon decay constants and the vector form factor of $K_{l3}$ \cite{Amoros:1999dp,Bijnens:2003uy,Bijnens:1998yu}.  The double logs amount to 
$20 - 35 \%$ of the
NNLO order result, corresponding to around $10\%$ of the total corrections at NNLO to the leading 
order result. Since the latter are of importance in the procedure of the chiral extrapolation 
discussed later, the double logs can make a sizeable contribution for such applications.
\begin{table*}[h]
\begin{center}
\begin{tabular}{ccccc}
 & LO & NLO & NNLO & Double Logs \\
 \hline
 $F_{\pi}/F_{0}$ & 1 & 0.068 & -0.172 & -0.050 \\
 $F_{K}/F_{\pi}$ & 1 & 0.216 & 0.035 & 0.06 \\
 $f_{+}(0)[K_{l3}]$ & 1 & -0.023 & 0.015 & 0.004 \\
\end{tabular}
\end{center}
\caption{The chiral corrections up to NNLO for the pion and Kaon decay constants and the
vector form factor of $K_{l3}$.
The values can however vary considerably, depending on the LEC's one employs. The numbers above are calculated with some standard values of the NLO LEC's and all
the renormalized NNLO LEC's set to zero at $\mu=770\MeV$.}
\label{t:b2l}
\end{table*}
In the framework of CHPT, the interactions between the mesons are encoded in the higher order contributions of the amplitude. So far these contributions can't be determined exactly,
since the low energy constants (LEC's) of the weak chiral Lagrangians at \ho 1 (NLO) are not known,
let alone the NNLO LEC's. The determination of the leading log contributions is thus the only possibility to get an estimate of the corrections one has to expect at NLO and NNLO.
In \cite{Laiho:2002jq} a method has been proposed to provide the LEC's needed for the \ktpp
amplitude from lattice simulations. \\
The outline of the paper is as follows: \\
In section \ref{s:cpt} I introduce some notation and the chiral Lagrangians used in the calculation.
Section \ref{s:ce} discusses the procedure  of the chiral extrapolation used in lattice
determinations of the \kztpp amplitude: The quantity which can be extracted from a lattice simulation
corresponds to the chiral limit of the amplitude, or, more accurately, the reduced counterpart thereof.
In a further step, the obtained chiral limit value has to be transformed back to the physical world with non-vanishing quark masses, which in the context of \cpt corresponds to the re-introduction of interactions
between the mesons (i.e. final state interactions), whose importance has already been pointed out. \\
This step and the numerics thereof are discussed in section \ref{s:nlocl} at NLO accuracy,
emphasizing the role of the chiral logs involved. In section \ref{s:dln} the procedure is extended
to include the double log contributions of this extrapolation, being part of the NNLO corrections to
the NLO results of section \ref{s:nlocl}. \\
In the appendices I outline the more technical aspects of the paper: \\
Appendix \ref{a:nlo} provides a thorough discussion of the NLO contributions and 
its approximation using only the chiral logs. 
Appendix \ref{s:cdlkpp} presents the procedure used to compute the double log contributions
of the \ktpp amplitude. The double logs are split into different classes of contribution:
The "genuine" double logs originating from one particle irreducible (1PI) two loop
topologies, which can be calculated with the NNLO counterterm Lagrangian provided in
\cite{Buchler:2005jk}. For a second class, the one particle reducible topologies, one can employ
separate one loop calculations for the respective 1PI subgraphs.
The results of the needed one loop expressions are
given in appendix \ref{s:ola}. The last contribution
originates from lower order diagrams where the masses and decay constants are shifted to their
physical value, as well as corrections coming from the wave-function
renormalization and LSZ procedure, up to the required chiral order.
The mass and decay shifts are provided in appendix
\ref{s:shifts}. Appendix \ref{s:oli} gives a very brief summary of the loop integrals used in
appendix \ref{s:ola}. \\
\section{\cpt Lagrangian}
\label{s:cpt}
Here I provide a very brief introduction of the Lagrangians which were used in 
the calculation. For more details about \cpt and in particular the definitions of the building blocks used throughout the paper I refer to the numerous review articles, for instance \cite{Ecker:1994gg,Colangelo:2000zw}. \\
The lowest order chiral Lagrangian which allows for $\De S =1 $ strangeness changing interactions is given by:
\be
 \mL^{(0)} = \mL^{(0)}_{\tiny{s}} + \mL^{(0)}_{\tiny{w}} \co
 \label{eq:slocl}
\ee
with the strong interaction Lagrangian:
\be
 \mL^{(0)}_{\tiny{s}} = \frac{F^{2}_{0}}{4} \big( \ltr \um\uum \rtr + \ltr \cp \rtr \big) \co
 \label{eq:loscl}
\ee
and the $\De S =1$ Lagrangian $\mL^{(0)}_{\tiny{w}}$:
\begin{alignat}{2}
 \mL^{(0)}_{\tiny{w}} & :=  C F_{0}^{4} g_{8} \ltr \De \um \uum \rtr \quad ; &
  \quad \De & := u \lambda_{6} u^{\dagger} \co
  \label{eq:lowcl}
\end{alignat}
where I introduce only the operator which dominates the contribution to the 
$I=0$ part of the amplitude. $\mL^{(0)}_{\tiny{w}}$ transforms like $(8,1)$ under $SU(3)_{L}\otimes SU(3)_{R}$. \\
In addition to the lowest order Lagrangians $\mL^{(0)}_{\tiny{s}}$ and $\mL^{(0)}_{\tiny{w}}$, I will also use the NLO Lagrangian $\mL^{(1)}_{\tiny{w}}$ \cite{Kambor:1990tz,Ecker:1993de}:
\be
 \mL^{(1)}_{\tiny{w}} = C F_{0}^{2} g_{8}  \sum_{i=1}^{37} N_{i}^{(1)} W_{i}^{(1)} \co
\ee
and the NNLO Lagrangian:
\bea
  \mL^{(2)}_{\tiny{w}} & = &  C g_{8}
                              \sum_{i=1}^{N} N_{i}^{(2)} W_{i}^{(2)} \co
  \label{eq:nnlowl}
\eea
with the respective coefficients:
\bea
N_{i}^{(1)} & = &
                    (\mu c)^{-\eps} \Big( N_{i}^{(1)\,r}(\mu,\eps) + Z_{i}^{1} \Lambda \Big) \co \\
 N_{i}^{(2)} & = &
 (\mu c)^{-2\eps} \Big( N_{i}^{(2)\,r}(\mu,\eps) +Z_{i}^{22}\Lambda^{2}
 + \big(Z_{i}^{21} + Z_{i\,L}^{21}(\mu,\eps)\big)\Lambda \Big)  \co
 \label{eq:nnlowct}
\eea
where $ Z_{i\,L}(\mu) $ is a coefficient originating from a vertex of $ \mL^{(1)} $. 
The notation is:
\be
 \Lambda = \frac{1}{(4\pi)^{2}} \frac{1}{\eps} \quad ; \quad \eps = \frac{1}{4 - d}\fs
\ee

\setcounter{equation}{0}

\section{Chiral extrapolation}
\label{s:ce}
For a lattice calculation of the \ktpp amplitude, one normally uses a tree
level PCAC relation to relate the former to the \ktp and \ktv amplitude
\cite{Bernard:1985wf}. The starting point is the lowest order Lagrangian, Eq.  \ref{eq:lowcl},
which implies the relation\footnote{The $\De$'s here shouldn't be confused with the chiral operator used in Eq. (\ref{eq:lowcl}).}:
\be
 \mA = \mA_{\mbox{\tiny{red.}}} + \mO(\hbar) \label{eq:ce} \co
\ee
with: 
\bea
 \mA & := & \ltr  \pi^{0} \pi^{0} | \mH_{\tiny{w}}^{(8,1)}  | K^{0} \rtr  \co \nn
  \mA_{\mbox{\tiny{red.}}} & := & 
-\i \frac{1}{F} \frac{ \De\phy}{m^{2} }
 \underbrace{ \ltr \pi^{0} | \mH_{\tiny{w}}^{(8,1)} | K^{0} \rtr }_{=: \mA_{\mbox{\tiny{red.}}}^{1}}  + \frac{1}{2} \frac{1}{F^{2}}
 \frac{\De\phy}{\De } \underbrace{ \ltr 0 |  \mH_{\tiny{w}}^{(8,1)} | K^{0} \rtr }_{=: \mA_{\mbox{\tiny{red.}}}^{2} } \label{eq:rme} \fs
\eea
I use the notation:
\begin{alignat}{2}
 \De  & :=   m_{K}^{2} - m_{\pi}^{2}
 \sem
 \quad  &  m^{2} & :=  p_{K}p_{\pi} \co
\end{alignat}
for the meson mass difference and intermediate meson masses which are 
evaluated on the lattice. $\De\phy$ corresponds to the physical value of $\De$. \\
Since this relation is derived from the tree level Lagrangian, it does not take into account any loop contributions,
in particular no re-scattering effects between the two pions (final state interactions).
However, these physical phenomena are reproduced in a lattice simulation, and consequently
one has to find a way to split a lattice result into a part corresponding to the tree level
and the higher order contributions. \\
To do so, we write Eq. (\ref{eq:ce}) again, noting that it is an exact identity if one
identifies all the matrix elements with the corresponding lowest order quantities
($X = \sum_{n=0}^{\infty} \hbar^{n} X^{(n)}$):
\be
 \mA^{(0)} = \mA_{\mbox{\tiny{red.}}}^{(0)} \co  
\ee
and re-introduces the higher order corrections with appropriate quotients:
\begin{alignat}{3}
\mA & := q \mA^{(0)} & \; ; \; 
\mA_{\mbox{\tiny{red.}}}^{1} & =  q_{1} \mA_{\mbox{\tiny{red.}}}^{1\,(0)}  & \; ; \; 
\mA_{\mbox{\tiny{red.}}}^{2} & =  q_{2} \mA_{\mbox{\tiny{red.}}}^{2\,(0)} \fs
\end{alignat}
With these definition one can write 
( $\ltr A_{\mu}^{X} \rtr = \ltr 0 | A_{\mu} | X \rtr = \i F p^{X}_{\mu} \, ; \, F = q_{F}F_{0}$ \footnote{I do not differentiate between $F_{\pi}$ and $F_{K}$ since they will only be relevant in their chiral limit where they coincide.}): 
\bea
 \mA & = & q \De\phy \Big( -\i F_{0} \frac{q_{F}^{2}}{q_{1}} \frac{\mA_{\mbox{\tiny{red.}}}^{1}}
 {\ltr A_{\mu}^{\pi} \rtr \ltr A_{\mu}^{K} \rtr} 
+ \frac{1}{2} \frac{q_{F}^{2}}{q_{2}} \frac{\mA_{\mbox{\tiny{red.}}}^{2}}
{\big( \ltr A_{\mu}^{K} \rtr^{2} - \ltr A_{\mu}^{\pi} \rtr^{2} \big) } \Big)
\label{eq:ce3} \fs
\eea
The expression in brackets on the RHS of Eq. (\ref{eq:ce3}) corresponds to the 
LEC $g_{8}$ and is 
thus independent of quark masses. Consequently one is free to
take the chiral limit $m_{q} \to 0$ of this quantity, which takes
$q_{1}$, $q_{2}$ and $q_{F}$ to their chiral values of 1: 
\bea
 \mA & = & q \De\phy \Big( -\i F_{0} \lim_{m_{q} \to 0} \frac{\mA_{\mbox{\tiny{red.}}}^{1}}
 {\ltr A_{\mu}^{\pi} \rtr \ltr A_{\mu}^{K} \rtr} 
+ \frac{1}{2} \lim_{m_{q} \to 0}\frac{\mA_{\mbox{\tiny{red.}}}^{2}}
{\big( \ltr A_{\mu}^{K} \rtr^{2} - \ltr A_{\mu}^{\pi} \rtr^{2} \big) } \Big)
 \fs
\label{eq:ce4}
\eea
One evaluates the RHS of Eq. (\ref{eq:ce3}) for various quark masses on the lattice and performs the extrapolation to the chiral limit, i.e. Eq. (\ref{eq:ce4}). 
Examples of how this procedure works in practice can be found in \cite{Noaki:2001un,Blum:2001xb}.  \\
The rest of this paper will be concerned about $q$:
\be
 q = \frac{\mA}{\mA^{(0)}} = 1 + \sum_{n=1}^{\infty} \hbar^{n} \delta q^{(n)} \co 
 \label{eq:q2def}
\ee
which is needed on the RHS of Eq. (\ref{eq:ce4}) in order to bring the chiral limit value 
of the \kztpp amplitude (i.e. $\De\phy $ times the expression in brackets) 
back to the real world with the physical (nonzero) quark masses. \\
In section \ref{s:nlocl} I discuss the NLO corrections to $q$, emphasizing the
contribution of the chiral logs.
Subsequently, in section \ref{s:dln}, I will include the double logs, which are a part
of $\delta q^{(2)}$, into the discussion. The calculation of these is outlined in appendix  \ref{s:cdlkpp}.

\setcounter{equation}{0}

\section{The chiral logs of \kztpp at NLO}
\label{s:nlocl}
Let me start by providing the NLO logarithmic corrections to the \kztpp amplitude to the lowest chiral order result. These corrections were first calculated by Bijnens \cite{Bijnens:1984qt}. \\
The \kztpp amplitude is expanded in its $\hbar$ order:
\be 
 \mA = \sum_{n=0}^{\infty} \hbar^{n} \mA^{(n)} \fs
\ee
At leading order, we have:
\be
 \mA^{(0)} = -\sqrt{2} \i C F_{0} g_{8} (\mk^{2} - \mpi^{2}) \fs
 \label{eq:loa}
\ee
Altogether six diagrams ( see Fig. \ref{f:nloa} ) contribute to the full NLO result, 
which is provided in Eq. (\ref{eq:ola}).
The $\nu$-dependence of $\mA^{(1)}$ is drawn as a full line in Fig. \ref{f:ab}.  \\
Restricting the NLO contributions to the chiral logs using
the approximations as outlined in appendix \ref{a:nlo}, one gets the expression:
\bea
 \mA^{(1)}_{\mbox{\tiny{log}}} & = &
- \sqrt{2} \i C F_{0} g_{8} \mk^{2}\Big( 1 - \mk^{2}\big(  L_{\pi}(\nu) + \frac{1}{4} L_{K}(\nu)
 \big)
  + \frac{27}{8} \mpi^{2} L_{\pi}(\nu) + ... \Big)  \co
  \label{eq:ceka1}
\eea
where I used the notation ($\nu$: renormalization scale):
\be
 L_{X}(\nu) := \frac{1}{(4\pi F)^{2}}\mln(\frac{m^{2}_{X}}{\nu^{2}}) \; ; \quad X = \pi,K,\eta \fs
 \label{eq:defL}
\ee
The approximation corresponding to Eq. (\ref{eq:ceka1}) is drawn as a dashed gray line in Fig. \ref{f:ab}, closest to the full result. The nearby
lightly dashed line corresponds to the leading order result plus the full logarithmic NLO contribution.

\begin{figure}[h]
\begin{center}
\includegraphics[width=8cm]{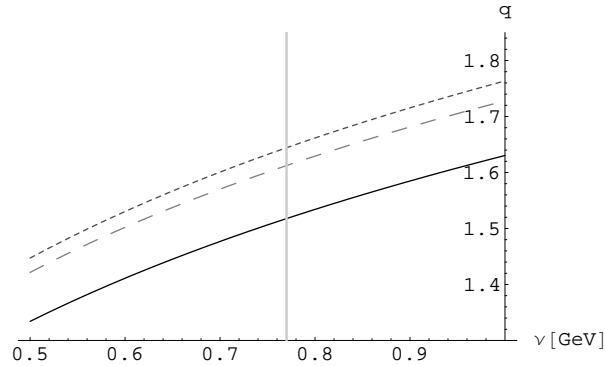}
\caption{\label{f:ab} The renormalization scale dependence of $q= |\mA^{(1)}|/\mA^{(0)} $ in terms of $F_{0}$. The full line corresponds to the full expression, Eq. (\ref{eq:ola}),  
with all LEC's $N_{i}^{(1)\,r}(\nu)$ set to zero. The lightly dashed gray line is the approximation taking at NLO only the (full) chiral logs into account.
The dashed gray line below, closest to the full result, corresponds to the approximation given in Eq. (\ref{eq:ceka1}).}
\end{center}
\end{figure}

\noindent In order to get an estimate for $q^{(1)} = |\mA^{(1)}|/\mA^{(0)}$ 
, I will use the natural scale of CHPT, $4\pi F_{\pi} / \sqrt{N_{f}} \simeq m_{\rho}$, $N_{f}$ being
the number of dynamical fermions \cite{Soldate:1989fh}, where one expects the values of the LEC's not to have acquired too large values due to running. Using $F_{0}$ as input,  one gets:
\bea
 q^{(1)}(\nu = 0.77 \GeV) & \simeq &  1.52 \;\; (1.61) \co
\eea
which corresponds to the full amplitude, as given in appendix \ref{s:ola}, with all NLO LEC's set
to zero; The number in brackets is the value of the approximation given in Eq. (\ref{eq:ceka1}).
Please note ( Fig. \ref{f:ab} ) that the latter simple expression is able to mimic the full amplitude
( with $N_{i}^{(1)\,r}(\nu) \equiv 0 $) with an accuracy well below $10\%$. Since the 
lack of knowledge of the LEC's introduces an error of the same order or probably even higher, it does not really make a big difference if one uses the simple form of  Eq. (\ref{eq:ceka1}) instead of the full expression, Eq. (\ref{eq:ola}).

\setcounter{equation}{0}

\section{Chiral extrapolation including the double logs}
\label{s:dln}
\begin{figure}[h]
\begin{center}
\includegraphics[width=8cm]{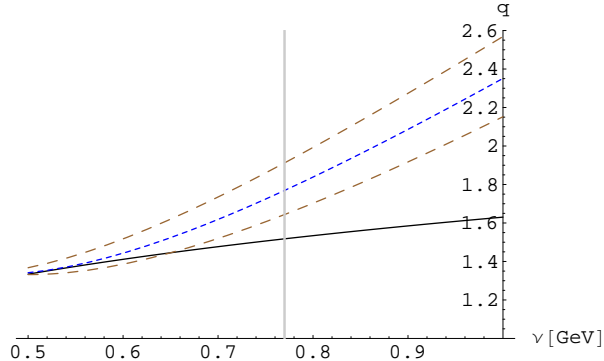}
\caption{\label{f:q2b} The renormalization scale dependence of $q$ at NLO (full line)
and at NLO with the
double log contributions included with an intermediate mass $m_{M}(t=0)$ (lowest dashed line), $m_{M}(t=0.05)$ ( intermediate lightly dashed line) and $m_{M}(t=0.10)$ (uppermost dashed line) respectively.
All $N_{i}^{r}(\nu)$'s are set to zero.
The most natural choice to get an estimate of $q$ is to set all LEC's to zero at the natural scale of CHPT, $\nu \simeq m_{\rho}$, which results in the values 1.52 (NLO), 1.64 $(t=0)$, 1.77 $(t=0.05)$ and
1.91 $(t=0.1)$  respectively.}
\end{center}
\end{figure}

\begin{figure}[h]
  \hfill
  \begin{minipage}[t]{.45\textwidth}
    \begin{center}
      \epsfig{file=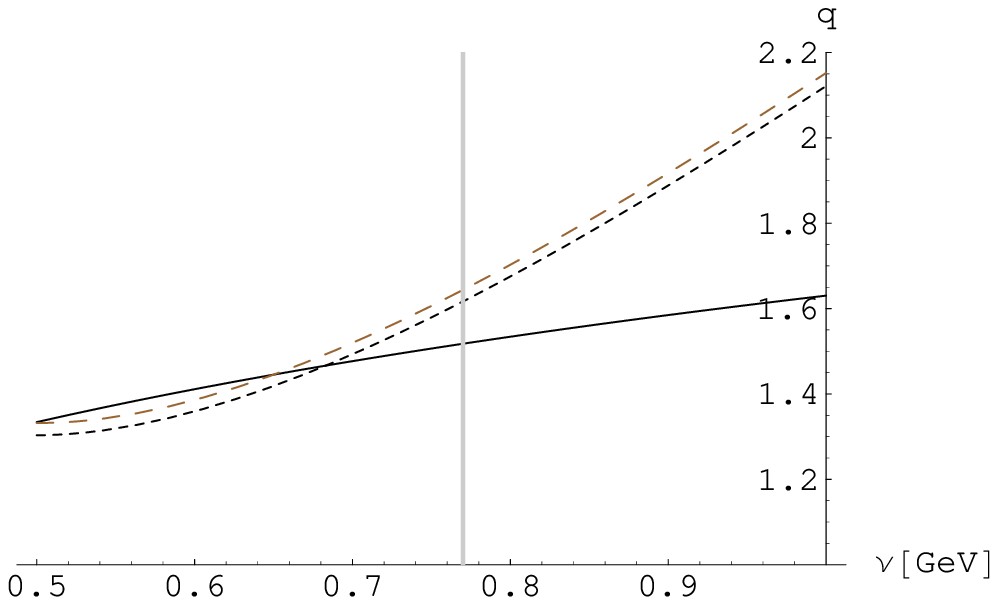, scale=0.5}
      \caption{The $\nu$ dependence of $q$ for $m_{M}(t = 0) = m_{K}$ using the full result of the
       double log contribution (lightly dashed line) and the approximation given in
       Eq. (\ref{eq:q2bf}) (upper dashed line). The full line correspond to the NLO result.}
      \label{f:q2dla}
    \end{center}
  \end{minipage}
  \hfill
  \begin{minipage}[t]{.45\textwidth}
    \begin{center}
      \epsfig{file=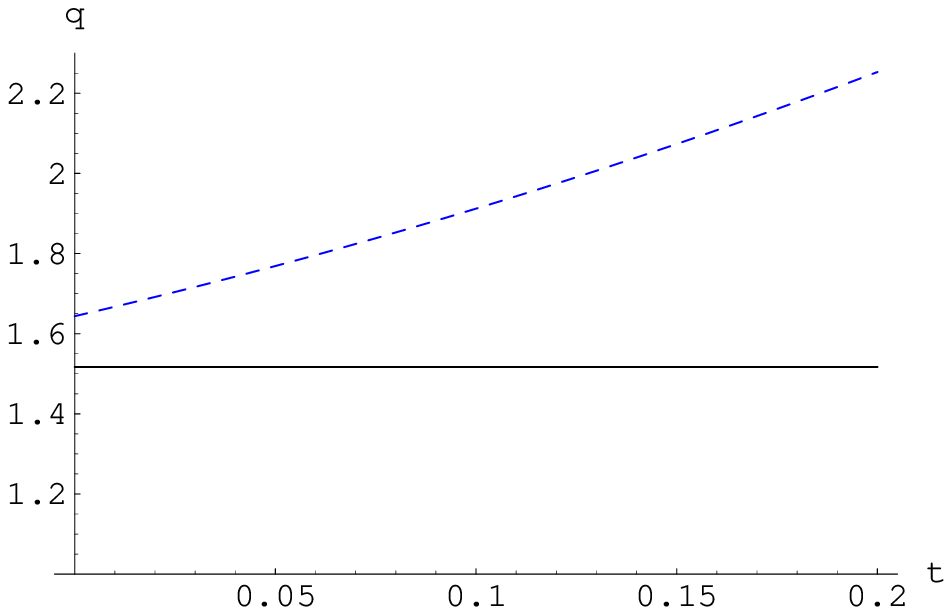, scale=0.5}
      \caption{The dependence on $q$ at $\nu = 0.77 \GeV$ on the meson mass value used in the
       logarithm $L_{M}$ in Eq. (\ref{eq:q2bf}), parametrized by $m_{M} =  \mpi^{t}\mk^{1-t}$. The
       full line corresponds to the NLO value.}
      \label{f:var}
    \end{center}
  \end{minipage}
  \hfill
\end{figure}

In this section I include the double log contributions to $\mA$  into the discussion of the factor $q$. The more technical details of how to calculate these is outlined in appendix \ref{s:cdlkpp}.
Using this the approximation, we have:
\bea
 \mA & \cong &
\mA^{(0)} + \delta\mA^{(1)} + \delta\mA^{(2)}_{\tiny{\mbox{log}}} \co \no \\
 & = & \mA^{(0)} \big( 1 + \delta^{(1)} + \delta^{(2)}_{\tiny{\mbox{log}}} \big) \fs
\eea
I will again work with the approximations for the chiral logs as outlined in appendix \ref{a:nlo},
i.e. Eq. (\ref{eq:kapr}) and (\ref{eq:logapr}). \\
In this scheme, the following terms contribute to the amplitude:
\bea
 \tilde{\delta}^{(2)}_{\tiny{\mbox{log}}}(\nu) & = & \mk^{4} \big(
 \tilde{\alpha}^{(2)}_{\pi K} L_{\pi}(\nu)L_{K}(\nu)  + \tilde{\alpha}^{(2)}_{K K} L_{K}(\nu)^{2} + \tilde{\alpha}^{(2)}_{M M} L_{M}(\nu)^{2}   \big)
 \co \label{eq:dla}
\eea
where I introduce an intermediate meson $m_{M}$ for the double log terms which were calculated via
the leading two-loop divergences. For these one cannot determine whether pions or 
kaons in the loops
generate the logs. The first two terms in Eq. (\ref{eq:dla}) originate from diagrams where we
performed the loop integration explicitly ( Terms proportional to $L_{\pi}^{2}$ are not generated).
In the following I will parametrize the intermediate mass M by:
\be
 m_{M}(t) = \mpi^{t}\mk^{1-t} \quad t \in [0,1] \no \fs
\ee
Similarly as for the NLO result, Eq. (\ref{eq:dla}) is not well defined in the limit where
$m_{\pi} \to 0 $ or $m_{M} \to 0 $. One should again view Eq. (\ref{eq:dla}) only to be an
allowed approximation for the double logs for the physical (nonzero) masses. \\
The term $\mk^{4} \tilde{\alpha}^{(2)}_{\pi K} L_{\pi}(\nu)L_{K}(\nu)$ in Eq. (\ref{eq:dla}) has a similar origin like the analogue piece of the NLO result, while the third term, 
$\mk^{4}\tilde{\alpha}^{(2)}_{M M} L_{M}(\nu)^{2}$, is generated by graphs with 
1PI two-loop subdiagrams. We will see below that the double log contribution is rather sensitive on the value chosen for $m_{M}$. The 1PI two-loop diagrams contributing to this term
are either of the "eight" or "sunrise" topology. For the former, the loop integrals are factorized, 
leading only to chiral logs of the form $m_{X}^{2}L_{X}$, from which follows that
$m_{M} = m_{K}$ or $t=0$ for this class of 
diagrams. For the sunrise topology, however, terms of the form $m_{K}^{4}\L_{\pi}^{2}$ and 
$m_{K}^{4}\L_{\pi}L_{K}$ can be generated
 \cite{Davydychev:1992mt,Amoros:1999dp}, presumably shifting $t$ to a nonzero value.
I am not aware of a method how to accurately estimate the value of $t$. 
However, it was already noted by Bijnens et al. in \cite{Bijnens:1998yu},  discussing the double logs in the strong sector,  that a large value of $t$ leads to unrealistic large double log contributions ( They used $t=1/2$ ).
In the following I will use the conservative values $t \in [ 0, 0.1 ]$.
The inability to estimate $t$ in an exact way unfortunately hinders us in making very accurate predictions for $q$. \\
The double logs originate from four classes of diagrams, as outlined in appendix \ref{s:cdlkpp}.
The separate contributions of these diagrams as well as more technical details can be found there.
Here I provide only the final result:
\bea
\mA^{(2)}_{\mbox{\tiny{log}}}(\nu) & = &
- \sqrt{2} \i C F_{0} g_{8} \mk^{2}\Big( 1 - \mk^{2}\big(  L_{\pi}(\nu)
  + \frac{1}{4} L_{K}(\nu)\big) + \frac{27}{8} \mpi^{2}L_{\pi}(\nu) \no \\
 & & + \mk^{4} \big( -  \frac{221}{108}L_{\pi}(\nu)L_{K}(\nu) - \frac{1349}{12960}L_{K}(\nu)^{2} + \frac{7703}{648} L_{M}(\nu)^{2}  \big)
+ ... \Big) \co
\label{eq:q2bf}
\eea
for the amplitude given in terms of $F_{0}$ \footnote{$F_{0}$ should also be used in the definition of the $L_{X}$, Eq. (\ref{eq:defL}).}. \\
Fig. \ref{f:q2b} shows the $\nu$ dependence of $q$ using the full NLO amplitude
($N_{i}^{r}(\nu) \equiv 0$) plus the double log contribution.
Fig. \ref{f:q2dla} illustrates the accuracy of the approximation, Eq. (\ref{eq:q2bf}), based on the use
of Eq. (\ref{eq:kapr}) and (\ref{eq:logapr}), by comparing it with $q$ calculated with  the "full"  result for the double logs. Fig. \ref{f:var} displays the dependence of $q_{2}$ on the intermediate
mass $m_{M}$ parametrized by $t$. \\
As already noted in \cite{Bijnens:1984qt}, the $\nu$-dependence of $q$ is quite strong.
This dependence would be canceled by the LEC's, whose values are unknown and consequently set
to zero at all values of $\nu$, making exact predictions impossible.
Employing the same reasoning as in section \ref{s:nlocl}, I will use the renormalization
scale $\nu =0.77 \GeV$ in order to get an estimate of $q$. Various values of $q$ at $\nu =0.77 \GeV$ at different values of $t$ are provided in Table \ref{t:qf}. \\
\begin{table*}[h]
\begin{center}
\begin{tabular}{c|c}
     & $q (\nu = 0.77 \GeV)$ \\
     \hline
  NLO & $1.52 \;\; (1.61)$ \\
  NLO + Double Logs $(t = 0.00)$ &  $1.64 \;\; (1.75)$ \\
  NLO + Double Logs $(t = 0.05)$ &  $1.77 \;\; (1.88)$ \\
  NLO + Double Logs $(t = 0.10)$ & $1.91 \;\; (2.03)$
\end{tabular}
\end{center}
\caption{The "enhancement factor" $q$ at NLO with and without the inclusion of the double logs at $\nu = 0.77 \GeV$, using $F_{0}$ as input. The numbers correspond to the full NLO expression provided in appendix \ref{s:ola}
with $N_{i}^{r}(\nu) \equiv 0$. The numbers in brackets correspond to the approximation as given in Eq.
(\ref{eq:q2bf}).}
\label{t:qf}
\end{table*}

\section{Impact on present lattice results}

In this section I discuss the implication of the presented results 
on the values for the 
 $\mbox{Re}(A_{0})$ provided from lattice simulations. \\
The experimental value of $\mbox{Re}(A_{0})$ is:
\be
 \mbox{Re}(A_{0})^{\tiny{\mbox{exp.}}} = 33.3 \cdot 10^{-8} \GeV
\ee
The CP-PACS collaboration calculated $\mbox{Re}(A_{0})$ using quenched domain wall fermions \cite{Noaki:2001un}. In their paper they use the reduced matrix 
element, Eq. (\ref{eq:rme}), and extrapolate it to the chiral limit, analoguesly to Eq. (\ref{eq:ce3}). For this means they either use a quadratic polynomial (Fit I: $\mbox{Re}(A_{0}) = \xi_{0} + \xi_{1}m_{M}^{2} + \xi_{2}(m_{M}^{2})^{2}$) or
chiral logarithms (Fit II: $\mbox{Re}(A_{0}) = \xi_{0} + \xi_{1}m_{M}^{2} + \xi_{2}m_{M}^{2}L_{M}$). The outcome is: 
\begin{displaymath}
 \mbox{Re}(A_{0})^{\mbox{\tiny{CP-PACS}}}_{\mbox{\tiny{chiral limit}}} = \left\{ 
\begin{array}{ll}
16.5 \cdot 10^{-8}\GeV & \textrm{Fit I} \\
20.7 \cdot 10^{-8}\GeV & \textrm{Fit II} 
\end{array}
\right.
\label{eq:cppacs}
\end{displaymath}
They do, however, not implement higher order \cpt corrections to the \ktpp amplitude, 
i.e. they use $q = 1$. 
If we take $q = 1.77$, corresponding to $\nu = 0.77\GeV$ and $t=0.05$ ( see section \ref{s:dln} ), one gets the values: 
\begin{displaymath}
 \mbox{Re}(A_{0})^{\mbox{\tiny{CP-PACS}}} = \left\{ 
\begin{array}{ll}
29.2 \cdot 10^{-8}\GeV & \textrm{Fit I} \\
36.6 \cdot 10^{-8}\GeV & \textrm{Fit II} 
\end{array}
\right.
\end{displaymath}
Since the above numbers can only serve as an order of magnitude estimate, I do not 
provide any error bars for them. The errors for $q$ are probably in the range of 
$20\%$ or even higher.  
The important statement one can however extract from the above
numbers is that the corrections due to $q$ can explain the discrepancy between 
the experimental number and the lattice value 
(which corresponds to the chiral limit value of $\mbox{Re}(A_{0})$ ). \\
The RBC collaboration \cite{Blum:2001xb} which also used quenched domain wall fermions in their simulations 
account for the chiral logs of the \ktpp amplitude, but use 
the old result of Bijnens \cite{Bijnens:1984qt}, which did not take into account all 
the diagrams contributing to these and is therefore incorrect
( see section \ref{a:nlo} ).
Their result with and without the "old" chiral logs is: 
\begin{displaymath}
\mbox{Re}(A_{0})^{\mbox{\tiny{RBC}}} = \left\{ 
\begin{array}{ll}
20.9 \cdot 10^{-8}\GeV & \textrm{chiral limit} \\
29.6 \cdot 10^{-8}\GeV & \textrm{with "old" chiral logs} 
\end{array}
\right.
\label{eq:rbc}
\end{displaymath}
If we again implement the factor $q = 1.77$, we get: 
\be
  \mbox{Re}(A_{0})^{\mbox{\tiny{RBC}}} = 37.0 \GeV \co
\ee
which agrees well with the experimental value within the rather large errors 
one has to assume. 
\section{Conclusion}
I have calculated the contributions of the (leading) chiral logs at NLO and at NNLO using Renormalization Group methods. I studied the result numerically and confirm and even reinforce the known result that
these contributions are quite large \cite{Bijnens:1984qt}.
With the approximations and assumptions used in this paper, the lowest order \ktpp amplitude is enhanced by a factor between $1.64$ and $1.91$ due to loop effects, depending rather strongly on the average value employed for the mass of the mesons in the loops generating the logs, but in all cases further enhancing the NLO value of $1.52$.
The numerical results provided here thus point in the right direction to explain the
$\De I = 1/2$ rule as a low energy QCD effect. \\
The above given numbers can only serve as a rough estimate of the higher order contributions,  since we are not in a position to include the contributions of the LEC's at NLO, and other additional contributions at NNLO.
However, given that it is not possible to calculate the higher order LEC's in this sector so far, these estimates are yet the best one can do with an acceptable effort to get an idea of the correction to the lowest order result one has to expect. \\
Even though not providing accurate numbers, the calculation presented here shows that one has to anticipate
rather large higher order \cpt corrections to the lowest order result of the $K \to (\pi\pi)_{I=0} $ amplitude.
This again indicates the importance to make further efforts to include these contributions, physically corresponding to the effects caused by the interaction between the three mesons, in an accurate way.

\section*{Acknowledgment}
I would like to thank Gilberto Colangelo for participation in the early 
stages of this work and many very helpful discussions. \\
This work was supported by the Swiss National Science Foundation.


\appendix
\renewcommand{\theequation}{\thesection.\arabic{equation}}
\setcounter{equation}{0}
\section{Detailed discussion of the NLO contributions}
\label{a:nlo}
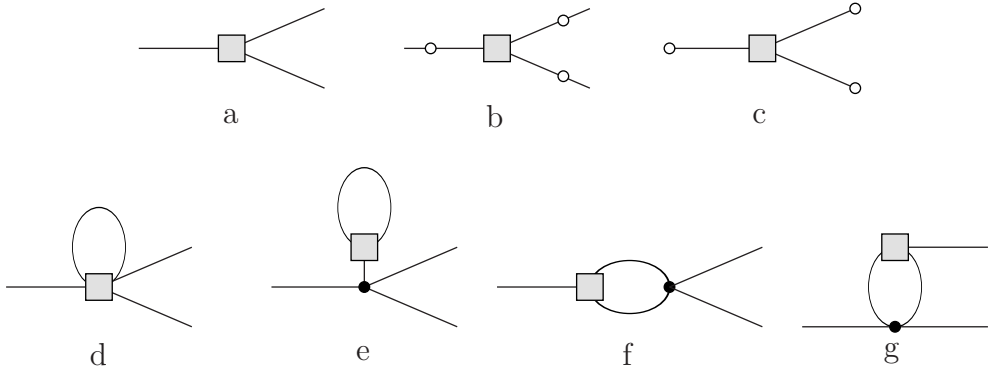
\begin{figure}[h]
\begin{center}
\begin{picture}(400,150)(0,0)

\Line(65,125)(100,125)
\Line(100,125)(135,110)
\Line(100,125)(135,140)
\GBoxc(100,125)(10,10){0.882}
\Text(100,100)[]{a}

\Line(165,125)(200,125)
\Line(200,125)(235,110)
\Line(200,125)(235,140)
\GBoxc(200,125)(10,10){0.882}
\GCirc(175,125){2}{1}
\GCirc(225,114.5){2}{1}
\GCirc(225,135.5){2}{1}
\Text(200,100)[]{b}

\Line(265,125)(300,125)
\Line(300,125)(335,110)
\Line(300,125)(335,140)
\GBoxc(300,125)(10,10){0.882}
\GCirc(265,125){2}{1}
\GCirc(335,140){2}{1}
\GCirc(335,110){2}{1}
\Text(300,100)[]{c}

\GOval(50,50)(15,10)(360){1}
\Line(15,35)(50,35)
\Line(50,35)(85,50)
\Line(50,35)(85,20)
\GBoxc(50,35)(10,10){0.882}
\Text(50,10)[]{d}

\GOval(150,65)(15,10)(360){1}
\Line(150,35)(150,50)
\Line(115,35)(150,35)
\Line(150,35)(185,50)
\Line(150,35)(185,20)
\GBoxc(150,50)(10,10){0.882}
\GCirc(150,35){2}{0}
\Text(150,10)[]{e}

\Line(200,35)(235,35)
\GOval(250,35)(10,15)(360){1}
\GBoxc(235,35)(10,10){0.882}
\GCirc(265,35){2}{0}
\Line(265,35)(300,20)
\Line(265,35)(300,50)
\Text(250,10)[]{f}

\Line(350,50)(385,50)
\GOval(350,35)(15,10)(360){1}
\GBoxc(350,50)(10,10){0.882}
\Line(315,20)(385,20)
\GCirc(350,20){2}{0}
\Text(350,10)[]{g}

\end{picture}
\end{center}
\caption{Diagrams contributing to the \kztpp amplitude at NLO: a: Leading order, b: Wave-function renormalization,
c: Decay constant renormalization, d: first tadpole, e: second tadpole, f: first unitary correction, g: second
unitary correction. Not drawn are the two counter-term diagrams.}
\label{f:nloa}
\end{figure}

In this section I give a detailed discussion of the NLO corrections and the 
approximations I made. The diagrams which contribute are drawn in 
Fig. \ref{f:nloa}. \\
\noindent Below I will work with the approximation where one restricts 
the NLO corrections to the chiral logs:
\bea
 \mA & = & \mA^{(0)} \big( 1 + \delta\mA^{(1)}_{\mbox{\tiny{log}}} + ... \big) \con
     & = & \mA^{(0)} \big( 1 + \alpha_{\pi}L_{\pi}(\nu)
     + \alpha_{K} L_{K}(\nu)
     + \alpha_{\eta} L_{\eta}(\nu) + ... \big)
     \fs
\eea
The $\alpha_{X}$ above are polynomials quadratic in the meson masses. \\
In addition I neglect all $\mpi^{2}$ contributions relative to $\mk^{2}$ in the polynomials ($\alpha$'s) and write the resulting quantities with a tilde:
\be
 \alpha_{X} = \mk^{2}\tilde{\alpha}_{X} + \mO(\mpi^2) \fs
 \label{eq:kapr}
\ee
This approximation will be particularly useful in the case of the NNLO contributions. \\
Furthermore I identify the logarithms:
\be
 L_{K}(\nu) \simeq L_{\eta}(\nu) \fs
\label{eq:logapr}
\ee
All following quantities for which the above approximations, Eqs. (\ref{eq:kapr}) and (\ref{eq:logapr}), are employed will be written with a tilde. \\
Using these, I get for $\delta\tilde{\mA}^{(1)}_{\mbox{\tiny{log}}}$:
\bea
 \delta\tilde{\mA}^{(1)}_{\mbox{\tiny{log}}}(\nu)
 & = &  \mk^{2} \big(  - L_{\pi}(\nu) - \frac{1}{4} L_{K}(\nu) \big) \fs
 \label{eq:damp}
\eea
This result corresponds to the lowest order amplitude in Eq. (\ref{eq:loa}), 
expressed in terms of the bare decay constant \footnote{Replacing $F_{0}$ with $F_{\pi}$
 shifts the constants in front of the chiral logs.}. 
In the following I will write all quantities in terms of $F_{0}$; Expressing 
amplitudes in terms of the renormalized decay constant $F_{\pi}$ increases generally the
$\nu$ dependence artificially. I use the values:
\begin{alignat}{2}
   F_{\pi} & =  0.0924 \; \GeV
 \sem
 \quad  &  \frac{F_{\pi}}{F_{0}} & =  1.06 \co
 \label{eq:fp0}
\end{alignat}
\noindent which should be precise up to two loop order \cite{Amoros:2001cp}. Using this approach, the results will
not get distorted by omitting known contributions to $F_{\pi}$ like LEC's etc. . \\ 
Please note that Eq. (\ref{eq:damp}) does not have a well defined limit $\mpi \to 0$. The above approximation is meant to be used with the physical values of the meson masses.
If we leave this regime and take the $\mpi \to 0$ limit, one has to include the associated $\bar{J}$ function in order to cancel the occurring divergence. 
In principle it would be aesthetically more appealing to have an expression at hand that respects the underlying chiral symmetries and limits,
which could be accomplished by re-introducing
parts of $\bar{J}$ contributions in Eq. (\ref{eq:damp}). 
I will however limit ourselves to
the strict use of the chiral logs as unambiguously defined in Eq. (\ref{eq:clo}), 
since this approach defines a well-defined setting.  \\
The separate contributions of the diagrams to $\tilde{\alpha}_{\pi}, \tilde{\alpha}_{K}$ and
$\tilde{\alpha}_{\eta}$ are given in Table \ref{t:ake}. \\
\begin{table*}[h]
\begin{center}
\begin{tabular}{cccccccc|c}
Diagram & a & b & c & d & e & f & g & total \\
\hline
$\tilde{\alpha}_{\pi}$ & 0 & 0 & 0 & 0 & 0 & -1 & 0 & -1 \\
$\tilde{\alpha}_{K}$ & $ 0 $ & 7/12 & 1/2 & -2 & 1/3 & - 1/6 & 1/2 & -1/4 \\
$\tilde{\alpha}_{\eta}$ & 0 & 1/6 & 0 & -22/27 & 8/27 & 0 & 23/27 & 1/2 \\
\hline
Sum  & 0 & 3/4 & 1/2 & -76/27 & 17/27 & -7/6 & 73/54 & -3/4 \\
\end{tabular}
\end{center}
\caption{The various contributions of the logs of the separate diagrams.}
\label{t:ake}
\end{table*}
In order to obtain an expression which agrees with the full logarithmic corrections at 
the $5 -10\%$ level, I also include the most dominant subleading contribution
\footnote{This is necessary since the coefficient of the subleading term is a factor 
3 times larger than the sum of the leading coefficients.}:
\bea
 \mA^{(1)}_{\mbox{\tiny{log}}} & = &
- \sqrt{2} \i C F_{0} g_{8} \mk^{2}\Big( 1 - \mk^{2}\big(  L_{\pi}(\nu) + \frac{1}{4} L_{K}(\nu)
 \big)
  + \frac{27}{8} \mpi^{2} L_{\pi}(\nu) + ... \Big)  \fs
  \label{eq:ceka11}
\eea

\noindent In the first paper treating these corrections the following value was provided \cite{Bijnens:1984qt}:
\be
 \tilde{\mA}_{\mbox{\tiny{log}}}^{\mbox{\tiny{B}}}(\nu)
\sim \frac{\mk^{2}}{F^{3}} \big( 1 - \frac{97}{27} \mk^{2} L_{K}(\nu) \big)
\fs
\label{eq:cekab}
\ee
Writing the result, Eq. (\ref{eq:damp}), with the above normalization, using
$L_{\pi} \simeq L_{K} \simeq \L_{M}$, in terms of $F_{0}$ and $F_{\pi}$
respectively, one gets:
\bea
\tilde{\mA}_{\mbox{\tiny{log}}} & \sim &
 \frac{\mk^{2}}{F_{0}^{3}}
  \Big( 1 - \frac{5}{4} m_{K}^{2}  L_{M}(\nu) \Big) \co
\label{eq:bep}
  \\
  & \sim &
\frac{\mk^{2}}{F_{\pi}^{3}}
  \Big( 1 - \frac{11}{4} m_{K}^{2}  L_{M}(\nu) \Big) \fs
  \label{eq:ceka111}
\eea
Equating $L_{\pi}$ and $L_{K}$ I found numerically to be a rather crude approximation;
I have checked that this identification results in values which disagree by approximately a factor of two from the full result of the logarithms. \\
In the original paper \cite{Bijnens:1984qt} the diagram e in Fig. \ref{f:nloa} is 
missing. This together with a different treatment of the $\mk^{2}L_{\pi}$ contributions
generated by diagram f is the origin of the above discrepancy \footnote{private communication with the author of \cite{Bijnens:1984qt}.}. 
Let us point out in this context that I checked my full NLO \ktpp amplitude 
on which the results above are based  with the output of the Mathematica package provided in \cite{Unterdorfer:2005au} and found
complete agreement.

\section{The calculation of the \ktpp double logs}
\label{s:cdlkpp}
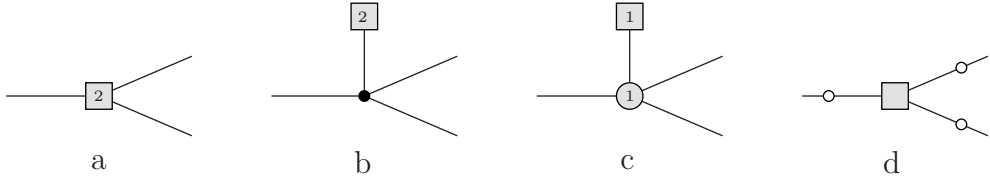
\begin{figure}[h]
\begin{center}
\begin{picture}(400,75)(0,0)

\Line(15,35)(50,35)
\Line(50,35)(85,50)
\Line(50,35)(85,20)
\GBoxc(50,35)(10,10){0.882}
\Text(50,35)[]{\tiny{2}}
\Text(50,10)[]{a}

\GBoxc(150,65)(10,10){0.882}
\Text(150,65)[]{\tiny{2}}
\Line(150,35)(150,60)
\Line(115,35)(150,35)
\Line(150,35)(185,50)
\Line(150,35)(185,20)
\GCirc(150,35){2}{0}
\Text(150,10)[]{b}

\GBoxc(250,65)(10,10){0.882}
\Text(251,65)[]{\tiny{1}}
\Line(250,40)(250,60)
\Line(215,35)(250,35)
\Line(250,35)(285,50)
\Line(250,35)(285,20)
\GCirc(250,35){5}{0.882}
\Text(251,35)[]{\tiny{1}}
\Text(250,10)[]{c}

\Line(315,35)(350,35)
\Line(350,35)(385,50)
\Line(350,35)(385,20)
\GBoxc(350,35)(10,10){0.882}
\GCirc(325,35){2}{1}
\GCirc(375,45.7){2}{1}
\GCirc(375,24.3){2}{1}
\Text(350,10)[]{d}

\end{picture}
\end{center}
\caption{Diagrams contributing to the \kztpp amplitude at NNLO. The square boxes represent 1PI loop graphs
with one insertion of a weak vertex and loop order given inside the boxes. The circles are the analogues
with no weak vertices. a: genuine 1PI, b: genuine 1PI subgraph attached to a strong tree vertex, c: A weak
and a strong subgraph with loop order 1 respectively, d: Lower order graphs (represented by a box with no associated loop order) which experience mass/decay constant shifts or LSZ renormalization. I have not drawn the associated counterterm diagrams. However, for diagram a and b I calculated the double logs via their
associated counterterms, and for diagram c and d I used partially the same approach.}
\label{f:dpc}
\end{figure}

\noindent In this section I provide the details of the calculation of the double logs of the
\ktpp amplitude:
\bea
 \mA & = & \la \pi(p_1) \pi(p_2) (I=0) | {\cal H}^{(8,1)}_{\tiny{w}}(q_w) |K(q_K) \ra \fs
\label{eq:gfdl}
\eea
Throughout the calculation, I allowed the weak Hamiltonian to carry the momentum $ q_w$,
a setting which can be used to account for the final state interactions
dispersively \cite{Buchler:2001nm}.
However, since the resulting expressions get too cumbersome, I will only provide
the results for the physical case, $q_{w} = 0$ \footnote{The full expression can
be obtained from the author.}.
As in the previous sections I will exclusively consider the octet
part of the nonleptonic weak chiral Lagrangian. \\
The double logs of \ho 2 contributing to Eq.
(\ref{eq:gfdl}) are produced by two-loop diagrams, each loop contributing
one logarithm.
The amplitude can be expanded in the \ho:
\be
 \mA = \mA^{(0)}+\mA^{(1)}+\mA^{(2)} + \mO(\hbar^{3})    \fs
\ee
All "genuine" double logs originating  from graphs with a two-loop 1PI
subgraph can be calculated with the $ \frac{1}{\eps^{2}} $ poles in
the two-loop counterterm, depicted in Fig.\ref{f:dpc}, diagrams a and b.
Additionally, there are graphs with two independent one loop subgraphs,
corresponding to diagram c. A further contribution originates from LSZ/wavefunction
renormalizations of graphs of lower chiral order, $\mO(p^{2},p^{4})$.
Finally, the lowest order masses and decay constants of these lower order
graphs have to be shifted to their renormalized physical value, up to the required
chiral order. This two last contributions are represented by diagram d. \\

\noindent For the graphs which consist of two one-loop subgraphs, diagram c, one can calculate the
subgraphs separately ( One-loop $K\pi$ scattering and $ K \to | 0 \rtr $ ).
However, there is a subtlety one has to account for: Since the Kaon which
propagates on the one-particle-reducible line of these graphs is offshell,
one cannot cut the graphs and identify the two resulting subgraphs as physical
processes, since for this to be true, the Kaon would have to be onshell
(an asymptotic state). As a consequence, one cannot treat the "genuine"
graphs a and b, the 1PR graph c,
and the LSZ/wavefunction contributions, diagram d, separately.
In particular, only the sum of these graphs will cancel all divergences. \\
For the graph c, I calculate the double logs explicitly by multiplying the two
separate one-loop graph expressions, and get a residual  $\frac{1}{\eps^{2}}$ divergence.
This divergence, however, can be
absorbed into the counterterm of the remaining, $\frac{1}{\eps^{2}}$ generating graphs. If we
proceed in this way,
the shifted counterterm will cancel the $\frac{1}{\eps^{2}}$ divergences of the rest of the
diagrams. \\
For those one can calculate the double logs as follows:
All loop divergences have the structure ( see Eq. (\ref{eq:clo}) ):
\bea
 \mathcal{Q} & = & \nu^{-\eps}
 \Big( \Lambda -\mu_{M}(\nu) + ... \Big)  \co
\eea
which ensures scale independence.
I used the definition: 
\be
 \mu_{X}(\nu) = \frac{1}{2} \frac{m_{X}^{2}}{F^{2}}L_{X}(\nu) \fs \no
\ee
If we define a basis $W_{i}^{(2)}$ of operators for the \ho 2 Lagrangian,
the sum of the divergences of all "genuine" two-loop diagrams will result in the
structure:
\bea
\label{eq:nnlost}
C g_{8} \sum_{i=1}^{N} \alpha_{i} W_{i}^{(2)} \mathcal{Q}^{2} & = &
  C g_{8} \nu^{-2\eps} \sum_{i=1}^{N} \alpha_{i} W_{i}^{(2)} \Big( \Lambda^{2}
 -2\Lambda\,\mu_{M}(\nu)
 + \mu_{M}(\nu)^{2} + ... \Big)  \fs
\eea
Note that one has generated a nonlocal divergence $\Lambda \,L_{M}$.
These structures will, however, be canceled by the contribution of one loop diagrams with
one insertion of a vertex from $\mL^{(1)}$.
Thus, the sum of the divergences of all diagrams will be free of nonlocalities and
be canceled by the Lagrangian $\mL^{(2)}_{\mbox{\tiny{w}}}$, Eq. (\ref{eq:nnlowl}). \\
Since the double logs of the loops always show up in the combination
$ \Lambda^{2} + \mu_{M}^{2}$, see Eq. (\ref{eq:nnlost}),
and the $ \Lambda^{2} $ divergences have to be canceled by the $ Z_{i}^{22}\Lambda^{2} $ term in
Eq. (\ref{eq:nnlowct}), the double logarithms can be calculated as follows
( The $\mfL_{i}^{22}$'s  correspond to the $1/\eps^{2}$ poles of the sum of all loop contribution,
$\mfL_{i}^{22} = -Z_{i}^{22}$):
\bea
\label{eq:gtl1}
 C g_{8} \nu^{-2\eps} \sum  \mfL_{i}^{22} W_{i}^{(2)} \mu_{M}(\nu)^{2} & = &
 - C g_{8}  \nu^{-2\eps} \sum Z_{i}^{22} W_{i}^{(2)} \mu_{M}(\nu)^{2}   \co \nn
 & = &  - C g_{8} \nu^{-2\eps} \mA^{2}_{2}\,\mu_{M}(\nu)^{2}  \fs
\eea
Note that of the chiral order $p^{2}$ and $p^{4}$ graphs, which are affected by wavefunction and
LSZ renormalization, only the tree graph will generate $\frac{1}{\eps^{2}}$ terms:
the one-loop graph is finite and the wavefunction and LSZ renormalization shifts on
the one-loop graphs are of \ho 1 and can therefore only introduce
$\frac{1}{\eps}$ divergences, and consequently we can discard the one-loop graph from the list of terms which
contribute to  $\frac{1}{\eps^{2}}$ divergences. \\
The final result of this calculation, split into a sum
of contributions originating from the different diagrams drawn in Fig. \ref{f:dpc}, is provided in
Table \ref{t:dlc}. \\
\begin{table*}[h]
\begin{center}
\begin{tabular}{cccc}
Diagram  & $\tilde{\alpha}^{(2)}_{\pi K}$ &
 $\tilde{\alpha}^{(2)}_{K K}$ &
 $\tilde{\alpha}^{(2)}_{M M}$  \\
\hline
 a & 0 & 0  & 18599/1944 \\
 b & 0 & 0  & -1213/972 \\
 c & -17/27 &  269/540 &  4/3 \\
 d & - 17/12 & -1561/2592 & 181/81 \\
\hline
Sum  & - 221/108 & - 1349/12960 & 7703/648 \\
\end{tabular}
\end{center}
\caption{The various contributions of the double logs of the separate diagrams, which are drawn in
Fig \ref{f:dpc}.}
\label{t:dlc}
\end{table*}

\setcounter{equation}{0}

\section{ One loop amplitudes}
\label{s:ola}
For completeness I provide here the full one loop \kztpp amplitude as well
as the \kkpp and \ktv amplitudes, needed in intermediate steps of the double log computation.
All these amplitudes have been calculated with the generalized kinematics, where the weak Hamiltonian is allowed to carry momentum. The results displayed below correspond however to the
amplitudes with the weak Hamiltonian at rest. The general result can be obtained from the authors. \\
The functions used in the amplitudes given below are defined in appendix \ref{s:oli}. \\
The renormalization scale dependence of the LEC's $L^{r}_{i}(\nu),N^{r}_{i}(\nu)$, which cancels the $\nu$-dependence of the chiral logarithms $\mu_{X}(\nu)$ will be suppressed in the following.
 All calculations have been performed with FORM \cite{Vermaseren:2000nd}.
\subsection{\kztpp}
\bea
\label{eq:ola}
 \lefteqn{\ltr \pi^{0} \pi^{0} | \mH_{\tiny{w}}^{(8,1)} | K^{0} \rtr
 = } \nn
 & &  \frac{i C F_{\pi} g_{8} \De}{\sqrt{2}}
  \Big( -2  + \frac{\hat{N}}{F^{2}} ( -\frac{8}{9} \mpi^{2} + 2\mk^{2} )
  + ( -\frac{27}{2} + 4\frac{\mk^{2}}{\mpi^{2}})\mu_{\pi}
     + \mu_{K}
     + ( \frac{37}{18} \frac{\mpi^{2}}{\me^{2}} -2\frac{\mk^{2}}{\me^{2}}) \mu_{\eta} \nn
 & &
      - \frac{1}{2}\frac{\mpi^{4}}{\De} (L_{\pi} - L_{\eta})
      + 4 \frac{\mpi^{2}}{F_{\pi}^{2}} (  8 L^{r}_{4} + 6 L^{r}_{5} -2 N^{r}_{5} - 4N^{r}_{7} - N^{r}_{8}
      +2N^{r}_{10} +4N^{r}_{11}+2N^{r}_{12}) \nn
 & &
     +\frac{4\mk^{2}}{F_{\pi}^{2}}(  16 L^{r}_{4} +  2 L^{r}_{5} - N^{r}_{5} +2 N^{r}_{7} - 2N^{r}_{8} - N^{r}_{9} )
     - \frac{2\mk^{2} -\mpi^{2}}{F_{\pi}^{2}} \bar{J}_{\pi\pi}(\mk^{2})
     - \frac{1}{9} \frac{\mpi^{2}}{F_{\pi}^{2}} \bar{J}_{\eta\eta}(\mk^{2}) \nn
 & &
     -  \frac{1}{\mpi^{2}F_{\pi}^{2}}\big( ( -\frac{1}{2} \mk^{4} +2 \mpi^{2} \mk^{2} )
      \bar{J}_{\pi K}(\mpi^{2})
      + \frac{1}{6} \mk^{4} \bar{J}_{K \eta}(\mpi^{2}) \big) \Big) + \mO(\hbar^{2}) \fs
\eea
I checked the above result with the Mathematica program recently provided
by \cite{Unterdorfer:2005au}.

 \subsection{\kkpp}
\bea
 \lefteqn{\ltr \pi \pi | \mH_{\tiny{s}} | K^{0}(q_{K}) \bar{K}^{0}(q_{\bar{K}}) \rtr_{q_{K}^{2} =  \mk^{2},\;q_{\bar{K}}^{2} =0 }
 =  }  \nn
  & &
 \frac{1}{2F_{\pi}^{2}} \Big( \frac{1}{3} \mk^{2}
  + \Lambda\big( \frac{13}{90}\mpi^{2} -\frac{17}{45}\mk^{2} \big)\mk^{2}
  + \frac{\hat{N}}{F^{2}} \big( \frac{1}{9}\mpi^{2}\mk^{2} -\frac{7}{12}\mk^{4} \big) \nn
 & &  + \mu_{\pi}  \big( 1 -\frac{\mk^{2}}{2\mpi^{2}} + \frac{\mk^{2}}{\De}(\frac{2}{3} -\frac{1}{6} \frac{\mk^{2}}{\mpi^{2}})  \big) \mk^{2}
 + \mu_{K} \big( \frac{11}{30} - \frac{1}{6} \frac{\mk^{2}}{\mpi^{2}} +\frac{1}{6} \frac{\mk^{4}}{\De\mpi^{2}} \big)\mk^{2} \nn
 & & \mu_{\eta} \big( \frac{1}{6} \mpi^{2} + \frac{1}{15} \mk^{2} + \frac{\mpi^{2}}{\me^{2}}(\frac{1}{18} \mpi^{2} -\frac{1}{3}\mk^{2} ) \big)
  + L^{r}_{4} \big( -\frac{56}{3}\mpi^{2}\mk^{2} + \frac{8}{3} \mk^{4} \big)
  - \frac{20}{3}L^{r}_{5} \mpi^{2}\mk^{2} \nn
  & & + 32 L^{r}_{6}\mpi^{2}\mk^{2} + L^{r}_{8}\big(\frac{8}{3} \mpi^{4} +32\mpi^{2}\mk^{2} + \frac{8}{3} \mk^{4} \big)
 + \bar{J}_{\pi}(\mk^{2}) \big( \frac{1}{6} \mpi^{2} - \frac{1}{3} \mk^{2} \big)\mk^{2} \nn
& &
 + \frac{1}{4}\bar{J}_{K}(\mk^{2})  \mk^{4}
 + \frac{1}{18} \bar{J}_{\eta}(\mk^{2})\mpi^{2}\mk^{2}
 + \bar{J}_{\pi K}(\mpi^{2})  \big( - \frac{1}{3} + \frac{1}{12} \frac{\mk^{2}}{\mpi^{2}} \big) \mk^{4}\Big)  + \mO(\hbar^{2}) \; .
\eea
 \subsection{\ktv}
 \bea
 \lefteqn{\ltr 0 | \mH_{\tiny{w}}^{(8,1)} | K^{0}(q_{K}) \rtr_{q_{K}^{2} = 0}
 = } \nn
 & & \frac{\i C F_{\pi}^{3}  g_{8}}{\sqrt{2}}
 \Big(  -\frac{20}{3} \De m_{K}^{2} \Lambda
       + 6 \mu_{\pi}m_{\pi}^2
       - 4 \mu_{K} m_{K}^2
       + \frac{2}{3}\mu_{\eta} \big(  m_{\pi}^2 - 4 m_{K}^2 \big) \nn
 &   & -8 \De \big(
       + 2N^{r}_{10} m_{K}^2
       + N^{r}_{11} (2 m_{K}^{2} - m_{\pi}^{2} )
       + 8  N^{r}_{23} m_{K}^2 \big) \Big)  + \mO(\hbar^{2}) \fs
\eea
Please note that the result given above correspond to the \kkpp and \ktv amplitudes evaluated at the unphysical point where the momentum of one of the Kaons vanishes. I checked that I agree with the results of \cite{Nehme:2001wf} for the \kkpp if I use the physical kinematics.

\setcounter{equation}{0}
\section{Renormalized masses and wavefunction renormalization}
 \label{s:shifts}
In subsection \ref{ss:m} I provide a list of the logarithmic contributions of the shifts of the bare masses to their renormalized value, up to NLO.
In subsection \ref{ss:wr},
the logarithmic contributions of the wavefunction renormalization shifts are given up to two-loop order.
\subsection{Masses}
\label{ss:m}
For the masses I only need the $\hbar = 1 $ corrections, since the tree Lagrangian given in Eq. \ref{eq:lowcl} generates only dynamical masses:
\be
 m_{X}^{2} = m_{X}^{(0)\;2} ( 1 + \delta m_{X}^{2} ) \; ; \quad X = \pi,K,\eta \fs
\ee
The log contributions of the NLO shifts thereof are \cite{Gasser:1985gg}:
\bea
 \delta m^{2\,(1)}_{\pi\,\mbox{\tiny{log}}} & = &  (\lp - \frac{1}{3} \le  )
\co \nn
\delta m^{2\,(1)}_{K\,\mbox{\tiny{log}}} & = & \frac{2}{3} \le  \con
\delta m^{2\,(1)}_{\eta\,\mbox{\tiny{log}}} & = & - \frac{\mpi^{2}}{\me^{2}} \le + ( 2 + \frac{2}{3} \frac{\mpi^{2}}{\me^{2}})\lk + ( \frac{7}{9}\frac{\mpi^{2}}{\me^{2}} -
	\frac{16}{9} \frac{\mk^{2}}{\me^{2}}\le ) \le \fs \no
\eea
\subsection{Wavefunction renormalization}
\label{ss:wr}
\be
 Z_{X} = 1 + \delta Z_{X} \; ; \quad X = \pi,K,\eta \co \no
\ee
with:
\be
 \delta Z_{X} = \sum_{n=1}^{\infty} \hbar^{n} \delta Z_{X}^{(n)} \fs \no
\ee
We split $\delta Z_{X\,\mbox{\tiny{log}}}^{(2)}$ into a piece linear
$(\delta_{1})$ and quadratic $(\delta_{2})$ in the logarithms:
\be
    \delta Z_{X\,\mbox{\tiny{log}}}^{(2)} = \delta_{1} Z_{X\,\mbox{\tiny{log}}}^{(2)}
                                 + \delta_{2} Z_{X\,\mbox{\tiny{log}}}^{(2)}
				 \fs \no
\ee
The one-loop contribution is \cite{Gasser:1985gg}:
\bea
 \delta Z_{\pi\,\mbox{\tiny{log}}}^{(1)} & = & \frac{4}{3}\lp + \frac{2}{3}\lk \co \nn
 \delta Z_{K\,\mbox{\tiny{log}}}^{(1)} & = & \frac{1}{2}\lp + \lk + \frac{1}{2} \le \con
 \delta Z_{\eta\,\mbox{\tiny{log}}}^{(1)} & = & 2\lk \fs \no
\eea
From $\delta Z_{X}^{(2)}$ one only needs $\delta_{2} Z_{X\,\mbox{\tiny{log}}}^{(2)}$ for the
calculation. Since the wavefunction renormalization at this order is not available in the literature
so far
, I use the counterterm Lagrangian provided in \cite{Bijnens:1999hw} to obtain it.
As already discussed in appendix \ref{s:cdlkpp}, 
one has to assume an unspecified intermediate meson mass $m_{M}$
in the loops generating the chiral logs: 
\bea
\delta_{2} Z_{\pi\,\mbox{\tiny{log}}}^{(2)} & = & \big( \frac{113}{36} \frac{\mpi^{4}}{m_{M}^{4}} -\frac{53}{18} \frac{\mpi^{2}\mk^{2}}{m_{M}^{4}} + \frac{73}{18} \frac{\mk^{4}}{m_{M}^{4}}\big) \mu_{M}^{2} \co \nn
\delta_{2} Z_{K\,\mbox{\tiny{log}}}^{(2)} & = &  \big( \frac{17}{9} \frac{\mpi^{4}}{m_{M}^{4}} -\frac{43}{36} \frac{\mpi^{2}\mk^{2}}{m_{M}^{4}} + \frac{32}{9} \frac{\mk^{4}}{m_{M}^{4}}\big) \mu_{M}^{2}
\con
\delta_{2} Z_{\eta\,\mbox{\tiny{log}}}^{(2)} & = & \big( \frac{5}{4} \frac{\mpi^{4}}{m_{M}^{4}}
 - \frac{1}{6} \frac{\mpi^{2}\mk^{2}}{m_{M}^{4}} + \frac{19}{6} \frac{\mk^{4}}{m_{M}^{4}} \big) \mu_{M}^{2}
\fs \no
\eea
The above expressions are not well-defined in the limit $m_{M} \to 0$, and one should only employ 
them for physical masses, i.e. $m_{M} > 0$ (See section \ref{s:nlocl}). 
\setcounter{equation}{0}
\section{Loop integrals}
\label{s:oli}
All tadpole diagrams translate into the integral:
\bea
 A_{a}  & := & \int \frac{d^{d}l}{(2\pi)^{d}}\frac{\i}{l^{2} - m^{2}_{a}} \co \\
	&  = & \frac{1}{(4\pi)^{d/2}}\Gamma(1 - \frac{d}{2})m^{\frac{d-2}{2}}_{a} \co \nn
	& = &  (\nu c)^{-\eps} m^{2}_{a} \big( -2\Lambda +          \frac{1}{(4\pi)^{2}}\mbox{ln}(\frac{m_{a}^{2}}{\nu^{2}})
	      \big) + \mO(\eps) \co \no
\eea
where I use:
\begin{alignat}{2}
 \eps & := 4 -d & \quad ; \quad \Lambda & := \hat{N} \eps  \co
\end{alignat}
 $\nu$ being the renormalization scale and $c$ parameterizing the renormalization prescription
( $c = -(\log(4\pi) - \gamma +1)/2$ in $\overline{MS}$).\
The simplest integral which occurs in the diagrams corresponding to the unitarity corrections is:
\bea
 J_{ab}(q^{2}) & := & -\i \int \frac{d^{d}l}{(2\pi)^{d}} \, \frac{\i}{l^{2} - m_{a}^{2}} \, \frac{\i}{(l-q)^{2} - m_{b}^{2}} \fs
\eea
Similar integrals to $J$ with a polynomial in $l$ in the numerator can be written in terms of a
combination of $A$ and $J$. $J$ can be split into a divergent and finite part as follows:
\bea
 J_{ab}(t) & = &  - \big( 2\Lambda + 2k_{ab} \big) + \bar{J}_{ab}(t)  \co
\eea
with
\be
 k_{ab} = \frac{\hat{N}}{2} \frac{\log(m_{a}^{2}) - \log(m_{b}^{2})}{m_{a}^{2} - m_{b}^{2}}
\ee
and
\bea
\bar{J}_{ab}(t) & = & J_{ab}(t) - J_{ab}(0) \co \\
		& = &  \hat{N} \Big(  1 + \big(  \frac{\De}{2t} - \frac{\Sigma}{2 \De}\big)
		\log(\frac{m_{a}^{2}}{m_{b}^{2}})
		- \frac{\las}{2t} \log(\frac{(t + \las)^{2} - \De^{2}}{(t - \las)^{2} - \De^{2}})
		       \Big)
		      \quad ; \lambda \geq 0  \co \nn
		& = &  \hat{N} \Big( 1 + \big(  \frac{\De}{2t} - \frac{\Sigma}{2 \De}\big)
		\log(\frac{m_{a}^{2}}{m_{b}^{2}}) \nn
		&   & - \frac{\las}{t}
		      \big( \arctan(\frac{t - \De}{\las}) - \arctan(\frac{-t - \De}{\las}) \big) \Big)
		      \quad \quad  ; \lambda <  0 \no \fs
\eea
with:
\begin{alignat}{2}
 \Sigma & := m_{a}^{2} + m_{b}^{2} & \; ; \quad \De := &
  m_{b}^{2} - m_{a}^{2} \fs \no
\end{alignat}
\begin{alignat}{2}
 \las & := \sqrt{ |\lambda|} & \; ; \quad \lambda := & \big(t-(m_{b}-m_{a})^{2}\big)\big(t-(m_{a}+m_{b})^{2}\big) \fs \no
 \end{alignat}
\bibliography{list}
\end{document}